\documentclass{article}

\usepackage{epsfig}

\def\be{\begin{equation}}

\def\ee{\end{equation}}

\def\bea{\begin{eqnarray}}

\def\eea{\end{eqnarray}}

\def\lsim{\mathrel{\lower4pt\hbox{$\sim$}}
\hskip-12.5pt\raise1.6pt\hbox{$<$}\;}

\def\gsim{\mathrel{\lower4pt\hbox{$\sim$}}
\hskip-12.5pt\raise1.6pt\hbox{$>$}\;}

\begin{document}
\newcommand{\preprintno}[1]
{\vspace{-2cm}{\normalsize\begin{flushright}#1\end{flushright}}\vspace{1cm}}

\title{\preprintno{{\bf ULB-TH/02-13}}Inflation from a Tachyon Fluid ?}

\author{Malcolm Fairbairn\thanks{mfairbai@ulb.ac.be}\\ Michel H.G. Tytgat\thanks{mtytgat@ulb.ac.be}\\
\\
Service de Physique Th\'eorique, CP225\\
Universit\'e Libre de Bruxelles\\
Bld du Triomphe, 1050 Brussels, Belgium}

\date{April 9, 2002}

\maketitle

\begin{abstract}
Motivated by recent works of Sen \cite{Sen:2002nu,Sen:2002in} and Gibbons \cite{Gibbons:2002md}, 
we study the evolution of a flat and homogeneous universe dominated by tachyon matter. In particular, we analyse the
necessary conditions for inflation in the early roll of a single tachyon field. 
\end{abstract}

\vfill \eject

\section{Introduction}

Despite numerous efforts, it seems difficult to reconcile string theory with the highly successfull paradigm of inflation. In
this brief note, following Gibbons \cite{Gibbons:2002md}, we investigate whether some form of Sen's tachyonic matter
\cite{Sen:2002nu,Sen:2002in} might provide the necessary ingredients for a phase of inflationary expansion in the early
universe.

According to Sen \cite{Sen:2002in}, (see also \cite{Gibbons:2001hf}), a rolling tachyon condensate in either bosonic or supersymmetric string theory can be
described by a fluid which in the homogeneous limit has energy density
\begin{equation}
\rho = {V(T)\over\sqrt{1 - \dot T^2}}
\end{equation}
and pressure
\be
\label{eos}
p = - V(T) \sqrt{1 - \dot T^2} \equiv - \rho(T) (1 - \dot T^2)
\ee
with $T$  the tachyon field and $V(T)$ the tachyon potential. 
These expressions are obtained from the tachyon matter effective lagrangian
\be
\label{lag}
{\cal L} = - V(T) \sqrt{1 - \dot T^2}.
\ee
Given the generic properties of $V(T)$ for $T\geq 0$,  a most remarkable feature of Sen's equation of state (\ref{eos}) is
that tachyon matter interpolates smoothly between 
\be
p = - \rho \;\;\Longleftrightarrow\;\; w = -1
\ee
for $\dot T=0$ and 
\be
p = 0 \;\;\Longleftrightarrow\;\; w = 0
\ee
as  $\dot T$ reaches its limiting value $\dot T=1$.  As already emphasized by Gibbons \cite{Gibbons:2002md}, if the tachyon
condensate starts to roll down the potential with small intial $\dot T$, a universe dominated by this new form of matter will
smoothly evolve from a phase of accelerated expansion to a phase dominated by a non-relativistic fluid. It is tempting to
speculate that the latter could contribute to some new form of dark matter. However, the topic of this paper is whether or not the tachyon condensate could  be relevant for inflation. (For related speculations, see for instance \cite{speculations}\cite{anupam}.) 

\bigskip

The  shape of the tachyon condensate effective potential depends on the system under consideration. In bosonic string theory for instance, this
potential has a maximum $V=V_0$ at $T=0$, where $V_0$ is the tension of some unstable bosonic D-brane, a local minimum with
$V=0$, generically at $T \rightarrow + \infty$, corresponding to a metastable closed bosonic string vacuum, and a runaway
behaviour for negative $T$. An exact classical potential ({\it{i.e.}} exact to all orders in $\alpha^\prime$, but only tree
level in $g_s$) encompassing these properties has been computed \cite{Kutasov:2000qp},
\be
\label{potential}
V(T) = V_0\left (1+ {T/T_0}\right)\exp{(-T/T_0)}.
\ee
Note that the curvature at the top of the potential (\ref{potential}) is 
$
{d^2V/ dT^2} = - {V_0/T_0^2}
$
. As the tachyon field has dimension $[T] = E^{-1}$, if $\dot T \ll 1$, from Eq.(\ref{lag}) we see that it is natural to
rescale $T$ by $\sqrt{V_0} T \equiv \phi$.  At $V=V_0$, the mass of the canonically normalized  $\phi$ is then $M_\phi^2 =
-1/T_0^2$. In \cite{Kutasov:2000qp}, $T_0 \sim l_s$ and $V_0$ is the tension of a bosonic Dp-brane, $V_0 \sim 1/g_s
l_s^{p+1}$.  
We shall see that these values of $V_0$ and $T_0$ seem marginally incompatible with inflation: the potential is simply too
steep. However, the values for $V_0$ and $T_0$ required to obtain slow-roll inflation are within an order of magnitude of
these values.

\section{Tachyon matter cosmology}

\begin{figure}
\begin{center}
\epsfig{file=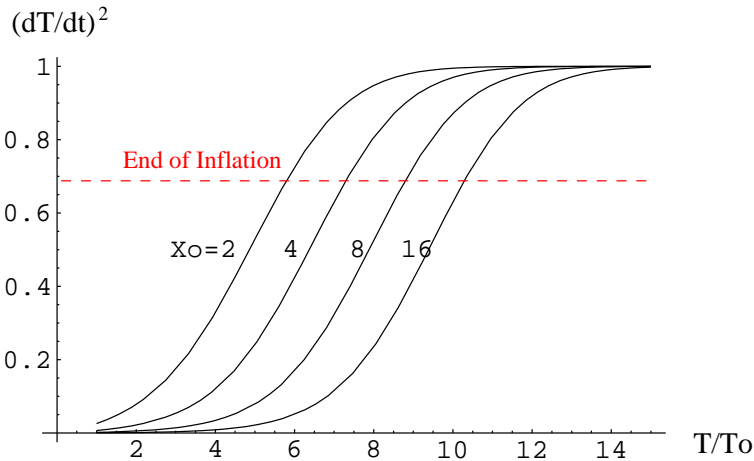,width=10cm}
\caption{\label{dotT2}$\dot T^2$ as a function of $T/T_0$, for various values of $X_0$. Inflation stops when $\dot T^2 >
2/3$.
}
\end{center}
\end{figure}

\begin{figure}
\begin{center}
\epsfig{file=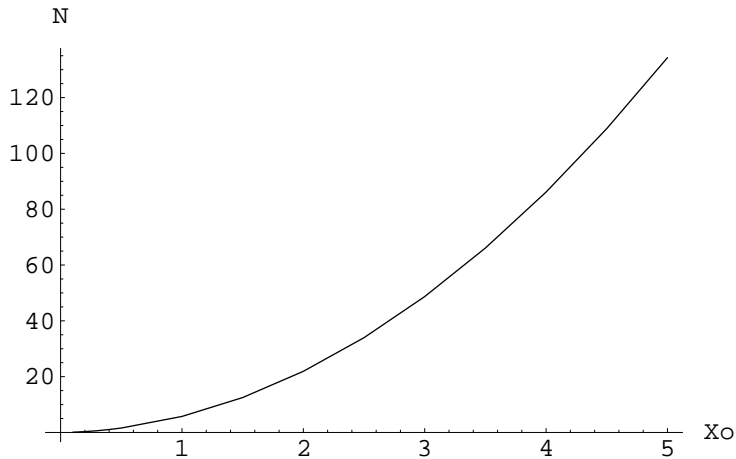,width=10cm}
\caption{\label{efolds} Number of e-folds during tachyon rolling as a function of $X_0$. 
}
\end{center}
\end{figure}

As shown by Gibbons, for a Roberston-Walker tachyon matter dominated universe, the Friedman equation takes the standard form
\be
H^2 = {\kappa^2 \over 3} \rho = {\kappa^2 \over 3}{V\over \sqrt{1 - \dot T^2}}
\ee
with $\kappa^2= 8 \pi G= 8 \pi /M_{pl}^2$ and where we have assumed spatial flatness\footnote{As tachyon matter turns into a
non-relativistic fluid for large $T$, a closed Universe will eventually recollapse. We assume flatness for simplicity.
Alternatively, the recollapse can be supposed to be in the far future of our model universe.} and have chosen to put the
cosmological constant to zero.
 Entropy conservation gives as usual
\be
\dot\rho = - 3 H (\rho + p)
\ee
the latter being equivalent to the equation of motion for the tachyon field $T$,
\be
\label{eom}
{V \ddot T\over 1 - \dot T^2} + 3 H\, V\,\dot T + V^\prime = 0
\ee
where $V^\prime = dV/dT$.
We would like to use these equations to determine the slow-roll conditions for inflation. The first condition to be satisfied
is that the expansion is accelerating (see for instance \cite{liddle} or \cite{Lyth:1998xn})
\begin{eqnarray}
 {\ddot a\over a} \equiv H^2 + \dot H &=& - {\kappa^2 \over 6}(\rho + 3 p) \;>\; 0\nonumber\\
&=& \frac{\kappa^2 }{3}{V\over \sqrt{1 - \dot T^2}}\left (1 - {3\over 2} \dot T^2\right)\;>\; 0
\end{eqnarray}
which requires that\footnote{This condition for inflation is in contrast to that obtained for a normal scalar field, $\dot \phi^2 < V(\phi)$.}
\be
\label{inflation}
\dot T^2 < {2\over 3}.
\ee
In order to have a sufficiently long period of inflation, the tachyon field should start rolling with small initial $\dot T$.
To relate the condition (\ref{inflation}) to the shape of the potential, we have to calculate the slow-roll conditons.
Following a standard procedure \cite{liddle}, we can express the evolution of the universe as a function of $T$ rather than
time. 
Using
\be
\dot H = - {\kappa^2 \over 2} {V \dot T^2\over \sqrt{1 - \dot T^2}} 
\ee
and because of the monoticity of $T$ with respect to time, we can rewrite this equation as
\be
H^\prime(T) = - {\kappa^2 \over 2} {V \dot T\over \sqrt{1 - \dot T^2}}
\ee
where the prime denotes derivation with respect to $T$.
Taking the square of the Friedmann equation we get 
\be
\label{HJ}
{H^\prime}^2  - {9\over 4} H^4(T) + {\kappa^4 \over 4} V(T)^2 = 0.
\ee
This first order differential equation is the analog for tachyon matter of the Hamilton-Jacobi form of the Friedmann equation
with a single inflaton field \cite{Salopek:1990jq}. It expresses the fact that $H^\prime$ is negligible as long as 
\be
H^2 \approx {\kappa^2 \over 3} V.
\ee
Solving (\ref{HJ}) for $H(T)$, we can then get $\dot T$ as a function of $T$
\be
{\dot T^2} = 1 - \left({\kappa^2 V\over 3 H^2}\right)^2.
\ee
Using the potential (\ref{potential}) one can solve (\ref{HJ}) numerically. We have chosen the field $T=T_i$ to be slightly
displaced from the maximum of the potential with initial $\dot T_i=0$. We also assume $T_i>0$ for obvious reasons.  The amount of inflation obtained depends upon the variable
$T/T_0$ and on a dimensionless parameter $X_0$ which characterizes the flatness of the potential close to its peak 
\be
X_0^2  = \kappa^2 T_0^2 V_0.
\ee
In Figure
\ref{dotT2} we plot $\dot T^2$ for different values of $X_0$. 
The pertinent feature of this figure is that for increasing $X_0$, inflation ends at increasing values of $T$. This
is equivalent to requiring that as the rolling of the field commences, the $\ddot T$ term is negligible in the equation of
motion for the tachyon field (\ref{eom}), so that
\be
3 H V \dot T \approx - V^\prime.
\label{slow}
\ee
Another quantity of interest is the number of e-folds during the inflationary phase,
\be
N(T) \equiv \ln {a(t_{end})\over a(t)} = \int_t^{t_{end}} H dt = - \int_T^{T_{end}} {H^2 V\over V^\prime} dT
\ee
This is shown in Figure \ref{efolds}. To successfully use inflation to solve the horizon problem and bring us a suitably flat
spectrum of perturbations we require $N~\geq~50-60$ e-folds of inflation whilst the field is  rolling.  As always, 
there is some uncertainty in the number of e-folds related to the choice of initial conditions. However, if we assume that $\dot T^2$ is small initially and that $T/T_0$ starts at less than about $0.1$ then the total number of e-folds becomes insensitive to the exact initial conditions. We choose the value of $X_0$ so as to accommodate enough e-folds and  the normalization to the COBE spectrum, which we shall describe in the next section. For the time being, we simply note that inflation lasts longer if $X_0$ is larger, as
shown in Figure \ref{dotT2}.
Once inflation is over the expansion of the tachyon-matter dominated universe slows rapidly which precludes the possibility of the modes generated during the slow roll regime being pushed outside our present day horizon.

\section{Slow-rolling tachyon matter}

The numerical analysis shows that inflation requires that $X_0 \gsim 3$. Once this is satisfied
the field $T$ evolves in a framework analogous to the slow-roll approximation for scalar fields. 
In order for the conditions
\be
H^2 \approx {\kappa^2\over 3} V
\ee
and
\be
H V \dot T \approx - V^\prime
\ee
to hold, the following inequalities must be satisfied:
\be
\label{epsilon}
\epsilon \approx {V^{\prime 2}\over \kappa^2 V^3} \ll 1
\ee
and
\be
\label{eta}
\eta \approx { V^{\prime 2}\over \kappa^2 V^3} - {V''\over \kappa^2 V^2} \ll 1.
\ee
These are just the usual definitions of the slow-roll parameters $\epsilon$ and $\eta$. Their form is different from the
usual expressions because we are taking derivatives with respect to $T$. Near $V_0$, for small $\dot T$, one can use the
canonically normalized field $\phi = \sqrt{V_0} T$ which brings (\ref{epsilon}) and (\ref{eta}) to their usual form
\cite{liddle}. As we assume that the tachyon is rolling from the top of the potential, generically $\epsilon \ll \eta$ for
small $T/T_0$, so that the second condition is the most stringent one.  Using the potential (\ref{potential}) to be
specific,  (\ref{eta}) becomes
\be
\kappa^2 V_0 T_0^2 \equiv X_0^2 \gg 1
\label{condition1}
\ee
a condition which is consistent with the numerical analysis  of the preceding section.

Finally we must estimate the size of the fluctuations and compare this estimate with the magnitude of the density
pertubations observed by COBE. 
Assuming that the slow-roll approximation can be made to hold over a significant range of cosmic scales, the spectral
index $n(k)$ is taken to be scale independent with
\be
n \approx 1 - 2 \eta \approx 1.
\label{spectrum}
\ee
We also follow the
standard procedure to estimate the density perturbations
\be
\left|\frac{\delta\rho}{\rho}\right|\approx\frac{H}{\dot{\phi}}\delta\phi\approx\frac{H}{\dot{T}}\delta T
\ee
where 
\be
{\sqrt{V_0} \delta T} \approx {H\over 2 \pi} \approx {\kappa \over 2 \pi} \sqrt{V_0\over 3}.
\ee
We then use the slow roll condition ($\ref{slow}$) to eliminate $V^{\prime}$ leading to the expression
\be
\left|\frac{\delta\rho}{\rho}\right|\approx\frac{\kappa^{3}V^{5/2}}{3^{3/2}2\pi\sqrt{V_0}V^{\prime}}
\ee
and we obtain the expression for the density perturbation during inflation when $T/T_0$ is of order one
\be
\left|\frac{\delta\rho}{\rho}\right|\approx\frac{1}{3^{3/2}2\pi}\frac{X_0^3}{\sqrt{V_0}T_0^2}
\ee
which correponds to roughly 60 efolds before the end of inflation for $5\lsim X_0 \lsim 7$. 
Since $\delta\rho/\rho \approx 2\times 10^{-5}$ we see that 
\be
\frac{X_0^3}{\sqrt{V_0}T_0^2} \equiv \kappa^{3} V_0 T_0 \approx 6\times 10^{-4}.
\label{condition2}
\ee
In order to calculate the magnitude of our perturbations, we assumed that $\dot{T}$ is negligible and that we are
close to the peak of the potential so that we can substitute $T$ for the canonical scalar field $\sqrt{V_0}\phi$.  However,
without a full derivation of the perturbations starting with the Lagrangian ($\ref{lag}$) we cannot say at what point our
approximations break down, and what kind of behaviour replaces the normal lore at that point.  

\section{Discussion}

We have not mentioned the precise origin of the tachyon field in question, but given the string theory origin of Sen's equation of state, it is tempting to write the parameters $V_0$ and $T_0$  in terms of the string length $l_s$ and the open string coupling constant $g_s$,
\be
V_0=\frac{v_0}{g_s l_s^4 (2\pi)^3}\qquad,\qquad T_0=\tau_0 l_s.
\ee
where $v_0$ and $\tau_0$ are dimensionless parameters such that $V_0/v_0$ is the tension of a D3-brane and $\tau_0 l_s$ is the inverse tachyon mass\cite{strings}.  The gravitational coupling in 4 dimensions is given in terms of the stringy parameters by
\begin{equation}
\kappa^2\equiv 8 \pi G_N =\pi g_s^2 l_s^2\left(\frac{l_s}{R}\right)^{6}.
\end{equation}
Here $R$ is the compactification radius of the compact 6 dimensional manifold, taken here to be a 6-torus.\footnote{Like in most higher dimensional scenarios, we assume that some unknown mechanism freezes the moduli associated to the extra dimensions. Also, for simplicity we assume that we can neglect the evolution of other fields, like the dilaton.} For the $D=4$ effective theory to be applicable one usually  requires that $ R \gg l_s$ since $R=l_s$ denotes the self T-dual point where the mass spectrum of KK and winding modes become degenerate \cite{strings}.  The volume of the compact space therefore must satisfy the inequality $V > (2\pi R)^6$.

In order for the gravitational waves at the end of inflation to be compatible with CMB observtions, there is a further condition $\cite{linde}$
\begin{equation}
\frac{H_{end}}{M_{pl}}\le 3.6\times 10^{-5}
\end{equation}
which together with the requirement for enough e-folds ($\ref{condition1}$) and the magnitude of the perturbations as given by equation ($\ref{condition2}$) leads to three conditions which can be written as
\begin{eqnarray}
\left(\frac{l_s}{R}\right)^{12}v_0 g_s^3 &\lsim& 1.4\times 10^{-6} \;\;\; \mbox{\rm (no grav. waves)}\label{ngw}\\
\left(\frac{l_s}{R}\right)^{9}v_0 \tau_0 g_s^2 &\sim& 2.7\times 10^{-2}\;\;\; \mbox{\rm(COBE normalization)}\\
\left(\frac{l_s}{R}\right)^{6}v_0 \tau_0^2 g_s &\gsim& 7.1\times 10^{2} \;\;\;\mbox{\rm (inflationary cond.)}\label{ic}
\end{eqnarray}
These conditions cannot all be satisfied simultaneously with the values one would choose as a first guess, $v_0 \sim \tau_0 \sim 1$, for any $R \gsim l_s$.\footnote{In the first version of this paper we overlooked a factor of $g_s$ in the expression of the Planck mass which led us to underestimate strong gravity effects.} 

For $R \sim l_s$ and insisting on $v_0 \sim 1$, (\ref{ngw}) gives $g_s \lsim 10^{-2}$ while (\ref{ic}) imposes  $\tau_0 \gsim 10^{2}$, corresponding to a rather light tachyon mass $m_T \lsim 10^{-2}/l_s$. This is phenomenologically the simplest solution, but is at odds with expectations from string theory. (See however \cite{Kim:2002rv}.)

If one insists on a tachyon mass $\sim 1/l_s$, the only alternative without entering the strong coupling regime is to increase the energy density, for instance by increasing the number of branes, in the spirit of \cite{anupam} where the potential arises from the combined tachyonic potential of a number of brane/anti-brane systems.  The problem here is that in order for each individual brane to act in the way described by Sen's effective action the typical seperation between branes within the compact directions would have to be larger than $l_s$.\footnote{If the branes are coincident, there are extra tachyonic degrees of freedom which all together transform either in the adjoint of the gauge group that lives on a stack of non-BPS D-brane or in the bi-fundamental for $Dp-\bar Dp$ pairs \cite{Horava:1998jy}.}
A miminal, although {\em ad hoc}, setting is to take $v_0$ to be equal to the number of string length size volumes within the compact space, i.e.
\begin{equation}
v_0=\left(\frac{2\pi R}{l_s}\right)^6
\end{equation}
Then it is possible to fulfill the above requirements with $g_s\sim 2\times 10^{-2}$, $R/l_s\sim 10$, corresponding to a very large number of branes ($\sim 10^{11}$).
This initial condition is quite baroque and at first sight rather unattractive. However we take note that such a large number is not foreign to speculations  on the dS/CFT correspondence \cite{Strominger:2001pn}, which suggests that the number of light degrees of freedom of the putative Euclidean CFT dual to the inflationary phase of the universe should be very large $c \sim 10^8$
\cite{Larsen:2002et}.

One last issue we would like to comment on is that of reheating at the end of inflation. In this paper our approach has been rather phenomenological. We took Sen's equation of state together with a string-theory motivated effective potential. Both ingredients rest on quite severe approximations, but are supposed to capture at least some of the physics of tachyon condensation. Remarkable features are that 1)  the minimum of the tachyon potential is at $T \rightarrow \infty$ and that 2) the energy of the brane stays confined to its hyperplane in the form of a pressureless fluid. As far as we understand, the first feature is supposed to be generic. As has been recently emphasized by Kovman and Linde \cite{linde}, this is quite worrisome for the purpose of reheating the universe at the end of inflation. They also note that this is potentially a problem for {\em all} string-inspired models of inflation based on the annihilation or decay of branes.  

From a phenomenological point of view the simplest resolution would be to have a  potential that vanishes at  finite $T$ so that reheating can proceed through  oscillations of  the  tachyon fluid around its minimum. Either way, this raises the issue of the decay of the tachyon fluid. The standard lore in the string-community is that unstable D-branes should ultimately decay into closed string modes. To make this process manifest one should take into account $g_s$ corrections to the tachyon effective action, a program which has not been completed yet. We believe that a related issue  will be to understand the precise nature of the tachyon fluid at the minimum of the potential. It is expected to be stable only in the limit $g_s \rightarrow 0$\cite{Sen} and should in principle be a good zeroth order approximation to the problem of tachyon matter decay. 

An intruiging characteristic of Sen's effective action is that at the minimum of the potential, the tachyon fluid equations of motions reduced to, in Hamiltonian form and in flat space-time, \cite{Sen:2002an}
\begin{eqnarray}
1 &=& (\dot T)^2 - (\partial_i T)^2\\
\dot \Pi &=& \partial_i \left(\Pi {\partial_i T\over \dot T}\right) 
\end{eqnarray}
where $\Pi$ is the momentum conjugate to $\dot T$.
These are precisely the equation of motion of a relativistic pressureless fluid of partons of mass unity, with velocity field $\vec v = \nabla T/\dot T \equiv \vec k/\omega(k)$ and energy density $\Pi(x,t)$. Solutions of these equations of motions are quite trivial to work out. A time-independent but otherwise arbitrary configuration of the energy-density $\Pi(x)$ corresponds in the parton picture to a bunch of massive partons {\em at rest}. However, generic solutions have caustic singularities and the free parton picture has probably a limited range.
It makes nevertheless clear why there are no propagating tachyon waves at the minimum of the potential, simply because there are no collective excitations in a pressureless fluid. 

In the limit of $g_s\rightarrow 0$, the tachyon fluid is stable and behaves as some new form of dark matter. The problem with this is that the universe would never become radiation dominated and for tachyon matter to be relevant in cosmology requires some fine-tuning \cite{Shiu:2002qe,linde}. For finite $g_s$ however tachyon matter should be unstable and decay. If we are willing to take the parton picture seriously, it is plausible that their decay could solve the issue of reheating of the universe at the end of tachyon rolling, however, not being string theorists, we have no idea how this question could be addressed. As an illustration we simply parametrize the decay rate
as
\begin{equation}
\Gamma \sim \frac{g_s^n}{l_s} \sim \left(\frac{l_s}{R}\right)^3 g_s^{n+1}M_{pl}
\end{equation}
The reheat temperature is then (presumably over-) estimated to be
\be
T_{RH}\sim\sqrt{M_{pl}\Gamma} \sim\left(\frac{l_s}{R}\right)^{\frac{3}{2}}g_s^{\frac{n+1}{2}}M_{pl}
\ee
which for generic powers of $n$ is high enough for most cosmological puposes. Obviously it would be interesting, both for string theory and for phenomenological applications to develop further insights on this issue.

\section{Conclusions}

In this paper we have investigated the possibility of using Sen's equation of state to see if we can obtain viable cosmological inflation.  We have described the conditions which the potential and the tachyon mass must fulfil in order to provide enough
e-folds of inflation and density perturbations of the correct magnitude. 
Our approach has been essentially phenomenological. In particular, the conditions for slow-roll are essentially independent of the precise shape of the potential and whether it vanishes at finite or infinite $T$. One attractive feature of Sen's equation of state is that it give an explicit realization of so-called k-inflation \cite{Armendariz-Picon:1999rj}. A related interesting property is that the exit from inflation is automatic for generic string-inspired effective potentials due to the limiting value $\dot T= 1$ for homogeneous configurations. As already emphasized numerous times a straightforward matching with string theory is problematic but perhaps not altogether impossible. An important issue, both for cosmological applications and, we presume, for string theory is to understand the ultimate fate of the tachyon matter at the bottom of the potential.

\section*{Acknowledgements}

We thank  Laurent Houart, Samuel Leach and Ashoke Sen for useful conversations or comments.

\end{document}